\documentclass[aps,prd,preprintnumbers,showpacs]{revtex4}
\usepackage[dvips]{graphicx}

\begin{document}

%
%

\eprint{Nisho-2-2020}
\title{Axion-Radiation Conversion by Super and Normal Conductors }
\author{Aiichi Iwazaki}
\affiliation{International Economics and Politics, Nishogakusha University,\\ 
6-16 3-bantyo Chiyoda Tokyo 102-8336, Japan.}   
\date{Sep. 24, 2020}
\begin{abstract}
We have proposed a method for the detection of dark matter axion. It uses superconductor
under strong magnetic field. As is well known, the dark matter axion induces oscillating electric field
under magnetic field. The electric field is proportional to the magnetic field and 
makes charged particles oscillate in conductors. Then, radiations of electromagnetic fields are produced.
Radiation flux depends on how large the electric field is induced and how large the number of
charged particles is present in the conductors.
We show that the electric field in superconductor is essentially identical to the one induced in vacuum.
It is proportional to the magnetic field. It is
only present in the surface because of 
Meissner effect.
On the other hand, although the magnetic field can penetrates the normal conductor,
the oscillating electric field is
only present in the surface of the conductor because of the skin effect. The strength of the electric field induced in the surface
is equal to the one in vacuum.
We obtain the electric field in the superconductor by 
solving equations of electromagnetic fields coupled with axion
and Cooper pair described by Ginzburg-Landau model.
The electric field in the normal conductor is obtained by solving equations of electromagnetic fields in the conductor
coupled with axion. 
We compare radiation flux from the cylindrical superconductor with that from the normal conductor with same
size. We find that the radiation flux from the superconductor is a hundred times larger than the 
flux from the normal conductor.  
We also show that when we use superconducting resonant cavity, we obtain radiation energy generated in the cavity 
two times of the order of the magnitude larger than that in normal conducting resonant cavity.
\end{abstract}
\hspace*{0.3cm}

\hspace*{1cm}

\maketitle

\section{introduction}
Axion is the Nambu-Goldstone boson\cite{axion} associated with Peccei-Quinn symmetry. 
The symmetry is a global U(1) chiral symmetry. It naturally solves the strong CP problem. Because the strong CP violation has not yet been observed,
we need to explain why the CP violating term $G_{\mu,\nu}\tilde{G}^{\mu,\nu}$ is
absent or extremely small in QCD Lagrangian where $G_{\mu,\nu}$ ( $\tilde{G}^{\mu,\nu}$ ) denotes ( dual ) fields strength of gluons.
The axion is a real scalar field which is the phase of a complex scalar field carrying the U(1) charge.
It acquires 
the mass $m_a$ through chiral anomaly. That is,
instantons in QCD give rise to the mass of the axion. The axion is called as QCD axion.
In the paper we mainly consider the QCD axion. The axion is generated in early universe 
and becomes one of dominant components of dark matter in the present universe when the Peccei-Quinn symmetry
is broken after inflation\cite{Wil}.

\vspace{0.1cm}
In the present day, the axion is one of the most promising candidates for the dark matter.
There proceed many projects for the detection of the axion. There are three types of the projects; Haloscope, Helioscope and others.
The dark matter axion in our galaxy is searched with Haloscope, while axion produced in the Sun is searched with Helioscope.
Haloscope projects are ADMX\cite{admx}, CARRACK\cite{carrack}, HAYSTAC\cite{haystac}, 
ABRACADABRA\cite{abracadabra}, ORGAN\cite{organ}, etc. Helioscope projects are CAST\cite{cast}, SUMICO\cite{sumico},  etc..
The others are LSW\cite{lsw} ( Light Shining through a Wall ), VMB\cite{vmb} ( Vacuum Magnetic Birefringence ) etc.
Since axion mass is unknown, the mass range in the search is very wide, $10^{-20}\rm eV$$\sim 1\rm keV$, or more\cite{obs}. 
There are axion-like particles proposed whose masses are not limited, differently to the QCD axion mass.
( The axion-like particles are those which couple with electromagnetic fields just as the QCD axion does. ) 
But we assume the QCD axion in this paper and expect that the appropriate mass range is of the order of $10^{-6}\rm eV$$\sim 10^{-4}\rm eV$,
It comes from cosmological consideration\cite{Wil} and simulation in latttice gauge theory\cite{lattice}. 

\vspace{0.1cm}
In previous paper\cite{iwa} 
we have proposed a method of the detection of radiations generated by dark matter axion. It is to use a superconductor of cylindrical shape.
In general, the dark matter axion induces oscillating electric field under strong magnetic field. 
The oscillating electric field makes Cooper pair oscillate in the superconductor so that the
radiations are emitted.
Similarly, we expect that radiations are also emitted by electrons in normal metal
under the magnetic field. In this paper we examine both cases in detail and show that 
the radiation flux from the metal is a hundred times smaller than that from the superconductor.

\vspace{0.1cm}
Our proposal is a type of Haloscope, in which the dark matter axion is converted to photon ( radio wave ) under magnetic field.
How large amount of radiation is generated depends on materials we use\cite{supp}. Here,
we consider superconductor and normal conductor ( sometimes we call it simply as metal ).
In the superconductor there are Cooper pairs typically with number density $\sim 10^{21}\rm cm^{-3}$, while 
there are electrons with number density $\sim 10^{22}\rm cm^{-3}$.
These charged particles emit radiations when they are forced to oscillate by the oscillating electric fields.
The strength of the electric fields are proportional to external magnetic field we impose.  
But, it is different depending on each material.

\vspace{0.1cm}
In this paper we show that the electric fields induced in the superconductor and the normal conductor
are essentially identical to the one induced in the vacuum. It is proportional to the magnetic field in the conductors.
However, the electric field in the superconductor is only present in the surface because
the magnetic field is only present in the surface to a penetration depth owing to the Meissner effect.
On the other hand, although the magnetic field penetrates the normal conductor, 
the oscillating electric field is only present in its surface because of the skin effect.
Theses oscillating electric fields generate oscillating electric currents.
The oscillating electric currents in both conductors
only flow through their surfaces and emit electromagnetic radiations.

To find how strong electric field is generated in the superconductor, we analyze Ginzburg-Landau model for the superconductor.
The superconductivity is represented as a condensed state of Cooper pair in the model.
We solve equations of electromagnetic fields coupled with axion and Cooper pair described by the model.
On the other hand, to find the electric field in the metal,
we solve equations of electromagnetic fields coupled with axion in the normal conductor
characterized by permeability $\mu$ and electric conductivity $\sigma$.

\vspace{0.1cm}
The presence of the oscillating electric fields leads to radiations from the conductors.
We show that the radiation flux from the superconductor is two times the order of
the magnitude larger than
that from the normal conductor. The difference arises from the difference between the penetration depth and the skin depth.
( The penetration depth in the superconductor is shorter than the skin depth as long as the frequency of the radiation
is approximately less than $100$GHz. ) The radiation flux from the superconductor is so large that existing radio telescope 
such as one with parabolic dish antenna of radius $\sim 10$m
can easily observe the radiation,
even when the superconductor is put at a hundred meter away from the telescope.
Indeed, the radiation flux from the cylindrical superconductor with radius $\sim 1$cm and length $\sim 10$cm 
under magnetic field $\sim 5$T is of the order of  $\sim 10^{-18}$W.

\vspace{0.1cm}
We also show that radiation energy generated in superconducting cavity is two times of the order of the magnitude larger than
that in normal conducting cavity, e.g. copper. The difference arises from the difference in the depth from the surface in which 
the oscillating electric current flows.

\vspace{0.1cm}
In the next section, we introduce axion photon coupling and find electric field in vacuum induced by axion under magnetic field.
In the section (\ref{3}), we introduce Ginzburg-Landau model for superconductor, in which Cooper pair is described by the field $\Phi$.
The model represents the coupling with electromagnetic field and the Cooper pair, and describes Meissner effect characterizing the superconductivity.
We derive electric field in the superconductor induced by axion under magnetic field.
In the section(\ref{4}) we solve the equations of electromagnetic fields in normal conductor coupled with axion.
We find that electric field in normal conductor induced by axion is only present in the surface of the conductor. 
In the section(\ref{5})
we numerically show the radiation fluxes from the superconductor and normal conductor of cylindrical shape. 
In the section(\ref{5.5}) we estimate radiation energy generated in superconducting resonant cavity. 
In the final section(\ref{6}) summary and discussion follow.

\section{electric field in vacuum}

First we show that the coherent axion induces an electric field in vacuum under a magnetic field.
It is well known that
the axion $a(\vec{x},t)$ couples with both electric $\vec{E}$ and magnetic fields $\vec{B}$ in the following,

\begin{equation}
\label{L}
L_{aEB}=k_a\alpha \frac{a(\vec{x},t)\vec{E}\cdot\vec{B}}{f_a\pi}\equiv g_{a\gamma\gamma} a(\vec{x},t)\vec{E}\cdot\vec{B}
\end{equation}
with the decay constant $f_a$ of the axion
and the fine structure constant $\alpha\simeq 1/137$,   
where the numerical constant $k_a$ depends on axion models; typically it is of the order of one.
The standard notation $g_{a\gamma\gamma}$ is such that $g_{a\gamma\gamma}=k_a\alpha/f_a\pi\simeq 0.14(m_a/\rm GeV^2)$
for DFSZ model\cite{dfsz} and $g_{a\gamma\gamma}\simeq -0.39(m_a/\rm GeV^2)$ for KSVZ model\cite{ksvz}.
In other words, $k_a\simeq 0.37$ for DFSZ and $k_a\simeq -0.96$ for KSVZ.
The axion decay constant $f_a$ is related with the axion mass $m_a$ in the QCD axion; $m_af_a\simeq 6\times 10^{-6}\rm eV\times 10^{12}$GeV.

\vspace{0.1cm}
We show that
the coupling parameter $k_a\alpha a/f_a\pi$ in the Lagrangian eq(\ref{L}) is extremely small for the dark matter axion $a(t)$.
Furthermore,
the dark matter axion $a(t)$ can be treated as a classical field because the axions are coherent.

We note that the energy density of the dark matter axion is given by

\begin{equation}
\rho_a= \frac{1}{2}(\dot{a}^2+(\vec{\partial} a)^2+m_a^2a^2)\simeq m_a^2a^2
\end{equation}
where $a(t)=a_0\cos(t\sqrt{m_a^2+p_a^2} )\simeq a_0\cos(m_a t)$, because the velocity $p_a/m_a$ of the axion is about $10^{-3}$ in our galaxy.
The local energy density of dark matter in our galaxy is supposed such as $\rho_a\simeq 0.3\rm GeV\,\rm cm^{-3}\simeq 2.4\times 10^{-42}\rm GeV^4$.
Assuming that the density is equal to that of the dark matter axion,
we find extremely small parameter $a/f_a\simeq \sqrt{\rho_a}/(m_af_a)\sim 10^{-19}$.
The energy density also gives the large number density of
the axions $\rho_a/m_a\sim 10^{15}\mbox{cm}^{-3}(10^{-6}\mbox{eV}/m_a)$, which causes their coherence. 
This allows us to treat the axions as the classical axion field $a(t)$. Anyway, 
we find that the parameter $k_a\alpha a(t)/f_a\pi=g_{a\gamma\gamma} a(t)$ is extremely small.

\vspace{0.2cm}
The interaction term in eq(\ref{L}) between axion and electromagnetic field slightly modifies Maxwell equations in vacuum,

\begin{eqnarray}
\vec{\partial}\cdot\vec{E}+g_{a\gamma\gamma}\vec{\partial}\cdot(a(\vec{x},t)\vec{B})&=0&, \quad 
\vec{\partial}\times \Big(\vec{B}-g_{a\gamma\gamma}a(\vec{x},t)\vec{E}\Big)-
\partial_t\Big(\vec{E}+g_{a\gamma\gamma}a(\vec{x},t)\vec{B}\Big)=0,  \nonumber  \\
\vec{\partial}\cdot\vec{B}&=0&, \quad \vec{\partial}\times \vec{E}+\partial_t \vec{B}=0.
\end{eqnarray}
From the equations, we approximately obtain the electric field $\vec{E}$
generated by the axion $a$ under the static background magnetic field $\vec{B}(\vec{x})$,

\begin{equation}
\label{E}
\vec{E}_a(r,t)=-g_{a\gamma\gamma}a(\vec{x},t)\vec{B}(\vec{x}),
\end{equation}
assuming the parameter $k_a\alpha a/f_a\pi$ extremely small.
This is the electric field in vacuum induced by the dark matter axion $a(t,\vec{x})=a_0\cos(\omega_a t-\vec{p}_a\cdot \vec{x})\simeq a_0\cos(m_at)$ 
with $\omega_a=\sqrt{m_a^2+p_a^2}\simeq m_a$.
We note out that the magnetic field configuration is arbitrary.

\section{electric field in superconductor}
\label{3}
Now we proceed to examine the electric field induced in superconductor.
Especially, we suppose that the superconductor is present at $x>0$ and uniform in $y$ and $z$ directions. The magnetic field
is parallel to the surface of the superconductor. It is also uniform in $y$ and $z$ directions, and points to $z$ direction.
We suppose that the superconductor is described by Ginzburg-Landau model with Cooper pair $\Phi$,

\begin{equation}
\label{gl}
L_{GL}=|(\partial_t-iqA_0)\Phi|^2-|(\vec{\partial}_x+iq\vec{A})\Phi|^2-h(|\Phi|^2-v_0^2)^2
\end{equation}
with electric charge $q=2e$  ( electron charge $e$ ) and coupling constant $h$, 
where $A_0$ and $\vec{A}$ denote gauge fields; $\vec{E}=-\vec{\partial} A_0-\partial_t\vec{A}$ and $\vec{B}=\vec{\partial}\times \vec{A}$.
We identify a field $\Psi$ used in nonrelativistic Ginzburg-Landau model such that $\Psi=\Phi \sqrt{2m_c}$ with mass $m_c$ of Cooper pair.
Because the number density $n_c$ of the Cooper pair is represented as $n_c=|\langle\Psi\rangle|^2$ in the nonrelativistic Ginzburg-Landau model, 
it is given such that $n_c=2m_c|\langle\Phi\rangle|^2=2m_cv_0^2$ in the relativistic Ginzburg-Landau model.  
( $m_c=2m_e$ with electron mass $m_e$. )
The superconducting state is represented by the condensed state $\langle \Phi \rangle=v_0$ of the Cooper pair.

\vspace{0.1cm}
When we include the effect of the Cooper pair, the modified Maxwell equations are led to the following,

\begin{eqnarray}
\label{M1}
\vec{\partial}\cdot\vec{E}+g_{a\gamma\gamma}\vec{\partial}\cdot(a(\vec{x},t)\vec{B})&+&2q^2A_0|\Phi|^2+iq\Phi^{\dagger}\partial_t\Phi+C.C. =0, \\ 
\label{M2}
-\vec{\partial}\times \Big(\vec{B}-g_{a\gamma\gamma} a(\vec{x},t)\vec{E}\Big)+
\partial_t\Big(\vec{E}+g_{a\gamma\gamma}a(\vec{x},t)\vec{B}\Big)&-&2q^2\vec{A}|\Phi|^2+iq\Phi^{\dagger}\vec{\partial}\Phi+C.C.=0,  \\
\vec{\partial}\cdot\vec{B}=0, \quad \vec{\partial}\times \vec{E}+\partial_t \vec{B}&=&0.
\label{M3}
\end{eqnarray}

In order to see magnetic field configuration in the superconductor, we solve the second equation(\ref{M2})
with external magnetic field $\vec{B}=(0,0,B_z(x))$ with $B_z(x)=B_0$ for $x<0$, neglecting the axion field,

\begin{equation}
\label{meissner}
-\vec{\partial}\times \vec{B}-2q^2\vec{A}_0|\Phi|^2=-\vec{\partial}\times \vec{B}-2q^2\vec{A}_0v_0^2=0 \quad \to \quad B_z(x)=B_0\exp(-x/\lambda)
\quad \mbox{for} \quad x \ge 0
\end{equation}
with $\Phi=v_0$ and the penetration depth $\lambda=(\sqrt{2}qv_0)^{-1}$. We have taken the value $\Phi=v_0$ of Cooper pair in the supercondoctor.
The equations (\ref{M1}) and (\ref{M3}) trivially hold when $A_0=0$ and $\Phi_0=v_0$.

We find the magnetic field configuration in the superconductor; it penetrates the surface to the depth $\lambda$. 
That is, it represents the Meissner effect.
Then, we expect that the electric field induced by the axion is also present only in the surface.

\vspace{0.1cm}
We will derive the electric field $\vec{E}_a$ induced by the axion under the magnetic field $\vec{B}(x)=(0,\,0, \,B_z(x))$.
Supposing the parameter $k_a\alpha a/f_a\pi$ extremely small and small momentum of axion $\vec{\partial}a=0$, 
we derive the equations for $\delta\Phi$ and $\delta\vec{A}$ from Ginzburg-Landau Lagrangian in eq(\ref{gl}) and 
modified Maxwell eq(\ref{M2}) 
($\Phi=\Phi_0+\delta\Phi$ and $\vec{A}=\vec{A_0}+\delta\vec{A}$ 
with $\vec{B}=\vec{\partial}\times \vec{A_0}$ ) ,

\begin{eqnarray}
\label{delta1}
0&=&-\partial_t^2\delta\Phi+(\vec{\partial}+iq\vec{A_0})^2\delta\Phi+2(\vec{\partial}+iq\vec{A_0})\cdot(-iq\delta\vec{A})v_0
-2hv_0^2\delta\Phi^{\dagger}+C.C  \\
0&=&(\vec{\partial}^2-\partial_t^2)\delta\vec{A}-2q^2v_0^2\delta\vec{A}+g_{a\gamma\gamma} \partial_t a\vec{B}
-2q^2\vec{A_0}v_0(\delta\Phi+\delta \Phi^{\dagger})-iqv_0\vec{\partial}\delta{\Phi}+C.C.
\label{delta2}
\end{eqnarray}
with the gauge condition $\vec{\partial}\cdot \delta\vec{A}=0$,
where $\delta\Phi$ and $\delta{\vec{A}}$ is of the order of $k_a\alpha a/f_a\pi$. 

In addition to eq(\ref{delta1}) and eq(\ref{delta2}), we have the equation(\ref{M1}) for the gauge field $A_0=\delta{A}_0$, 

\begin{equation}
\label{11}
0=\vec{\partial}\cdot (\vec{E}_a+g_{a\gamma\gamma}a\vec{B})+2q^2\delta{A}_0v_0^2+iqv_0\partial_t\delta\Phi+C.C. .
\end{equation}

All of the variables $\delta{\vec{A}}$, $\delta A_0$, and $\delta\Phi$ may oscillate according to the oscillation $a(t)\propto \cos(m_at)$.
Because we assume that the fields are uniform in $y$ and $z$ directions, we simplify the equation(\ref{delta2}) by taking
$\delta{\vec{A}}=(0,0, \delta{A}_z)$  as well as $\vec{A}_0=(0, A_{0y}, 0)$ and $\delta A_0=0$  \big( $\vec{E}_a=(0,0,-\partial_t\delta A_z) $ and $\vec{B}=(0,0,\partial_x A_{0y})$ \big),

\begin{equation}
\label{delta3}
0=(\vec{\partial}^2-\partial_t^2)\delta A_z-2q^2v_0^2\delta A_z+g_{a\gamma\gamma}\partial_t a\vec{B},
\end{equation}

It is easy to obtain
a solution of eq(\ref{delta1}), eq(\ref{11}) and eq(\ref{delta3}),

\begin{equation}
\delta\Phi=0, \quad \delta{A}_z=\big(A'_0\cos(\omega t )+A''_0\sin(\omega t)\big)\exp(-\frac{x}{\lambda'})   
 -\frac{g_{a\gamma\gamma} \partial_t a}{m_a^2}B_z,
\end{equation}
with 
arbitrary amplitude $A'_0$, $A''_0$ and frequency $\omega$, 
where we have $\lambda' \simeq \lambda(1+\omega^2\lambda^2/2)$, assuming
$\omega^2 \ll \lambda^{-2}=2q^2v_0^2$. 

Namely, the electric field is

\begin{equation}
E_{a,z}=\omega \big(A'_0\sin(\omega t)-A''_0\cos(\omega t)\big)\exp(-\frac{x}{\lambda})-g_{a\gamma\gamma} a(t)\vec{B} 
\end{equation}
with $\vec{E}_a=(0,0,-\partial_t A_z)=(0,0,E_{a,z})$

The first term in $E_{a,z}$ represent a radiation with frequency $\omega$ incoming from outside the superconductor, while the second term represents
the electric field $\vec{E}_a=-g_{a\gamma\gamma} a(t)\vec{B}(x)$ induced by the dark matter axion.
It is just equal to the one in eq(\ref{E}) in the vacuum.
In general,
the radiation incoming the superconductor from outside 
only penetrates the surface of the superconductor to the penetration depth $\lambda'\simeq \lambda$
owing to the Meissner effect. We should note that the radiation ( electric field ) in the outside is produced by the axion 
owing to the coupling between electromagnetic fields and axion in the presence of external magnetic field $B$.
It is just $\vec{E}_a=-g_{a\gamma\gamma} a(t)\vec{B}$. The field determines the first term of the solution $E_{a,z}$.
Thus, $\omega=m_a$ $A'_0=0$ and $A''_0=g_{a\gamma\gamma}a_0B/m_a$. 

It is instructive to see that we have small suppression factor $(m_a\lambda)^2$ in the electric field $\vec{E}_a$ 
when the second term $2q^2v_0^2\delta{\vec{A}}=\lambda^{-2}\delta{\vec{A}}$ in eq(\ref{delta3}) is absent. 
Namely, $\vec{E}_a=-(\lambda m_a)^2g_{a\gamma\gamma} a(t)\vec{B}$. ( Typically , $\lambda\sim 10^{-6}$ cm and $m_a^{-1}\sim 10$cm. )
The second term represents the effect of the Cooper pair and
cancels the derivative $\partial_x^2\delta{\vec{A}}$
in eq(\ref{delta3}) because $B_z\propto \exp(-x/\lambda)$. Thus, we find that the suppression factor is absent in the superconductor; 
$\vec{E}_a(x,t)=-g_{a\gamma\gamma} a(t)\vec{B}(x)$. The electric field $\vec{E}_a$ is only present in the surface of the superconductor because
the magnetic field $\vec{B}$ is only present in the surface. These results were used in the previous paper\cite{iwa}. 

\vspace{0.1cm}
Here we make a brief comment that similar electric field $\vec{E}_a=-g_{a\gamma\gamma} a(t)\vec{B}$ is induced
around magnetic vortices in type 2 superconductor.
As we know, the magnetic field penetrates the type 2 superconductor and forms vortices.
They are described as classical solutions in the Ginzburg-Landau model.
A solution of the magnetic vortex located at $\rho=0$ is characterized in the following,

\begin{eqnarray}
\Phi_v&=&|\Phi_v(\rho)|\exp(in\theta) \quad \mbox{with} \quad |\Phi_v(\rho=0)|=0 \quad \mbox{and} \quad |\Phi_v(\rho)|-v_0 \propto \exp(-\rho/\xi) 
\quad \mbox{for} \quad \rho \gg \xi \\
\vec{B}_v&=&(0,0,B^v_z(\rho)) \quad \mbox{with} \quad B^v_z(\rho)\propto \exp(-\rho/\lambda) \quad \mbox{for} \quad \rho \gg \lambda 
\end{eqnarray}
with integer $n$,
where the cylindrical coordinate $(\rho,\theta,z)$ is used. 
The important feature of the vortex is the flux quantization; $\int d\rho d\theta B^v_z=2n\pi/q$ with integer $n$.
The type 2 superconductor has a property that the penetration depth $\lambda$ is larger than the coherent length $\xi$ of the Cooper pair; $\lambda >\xi$.
Thus, we may approximate these fields such as $|\Phi_v(\rho)|=v_0$ and $B^v_z\propto \exp(-\rho/\lambda)$ for $\rho>\lambda$.
In the region $\rho>\lambda$, we have a similar equation to eq(\ref{delta3}) for $\delta \vec{A}\equiv \vec{A}-\vec{A}_0=(0,0,\delta A_z)$ in the cylindrical coordinate
by putting $\delta \Phi=\delta A_0=0$ and $\vec{A}_0=(0,A_{\theta},0)$; $B^v_z(\rho)=\partial_{\rho}A_{\theta}+A_{\theta}/\rho$ and $\delta\Phi\equiv \Phi-\Phi_v$. 
Then, we find that the solution of the electric field $\vec{E}_a=-\partial_t\delta\vec{A}$ is given such as
$\vec{E}_a(\rho)=-g_{a\gamma\gamma} a\vec{B}_v(\rho)$. The solution holds only in the region ( $\rho >\xi$ ) outside the vortex core.
This is the electric field expected naively when the magnetic field $B_v(\rho)$ is present.
Obviously, it is attached to the vortex.

\vspace{0.1cm}
We would like to mention that the current density $J_c$ induced by the axion is derived from eq(\ref{M2}) 
such as $J_c=2q^2v_0^2\delta A_z=\lambda^{-2}\delta A_z
=-\lambda^{-2}\int^t dt' E_a(t')=\lambda^{-2}\partial_t E_a(t)/m_a^2$ in the superconductor; $E_a\propto \cos(m_at)$. 
The current is superconducting current given by the Ginzburg-Landau model. Actually,
the current $J_c\propto \sin(m_at)$ has no dissipation of its energy; $\int_0^{2\pi/m_a} dt J_c(t)E_a(t)=0$.
The current is a superconducting current even in the presence of the electric field. 
The fact is valid both for the surface current in
the superconductor and the current flowing along the magnetic vortex.   

\vspace{0.1cm}
We note that the penetration depth is rewritten such as $\lambda=\sqrt{1/2q^2v_0^2}=\sqrt{m_c/q^2n_c}$ in terms of the number density $n_c$,
where we use the relation $n_c=2m_cv_0^2$ mentioned above. Then,
the current density is also rewritten such as $J_c=-\lambda^{-2}\int^t dt' E_a(t')\simeq q^2n_c E_a/m_a m_c$.  We will explain later
that the formula
can be also obtained by using the Drude model. The model describes the classical motion of 
Cooper pair under the electric field $E_a$.
  
\vspace{0.1cm}
In this section,
we find that the electric field induced in the superconductor is essentially identical to the one induced in the vacuum, although
the electric field is confined to the surface of the superconductor. It oscillates with the frequency identical to the one of the axion field $a(t)\propto \cos(\omega_at)$.
Similarly, the current density $J_c$ also oscillates with the same frequency.

\section{electric field in normal conductor}
\label{4}

Next, we derive the electric field in the normal conductor ( metal ) with permeability $\mu$ and electric conductivity $\sigma$. 
The configuration of the metal is the same as the case of the superconductor. That is, the metal is present at $x>0$ and uniform in $y$ and $z$ directions.
We impose magnetic field $\vec{B}=(0,0,B_z)$ parallel to the metal. The magnetic field can penetrate the metal. We denote the magnetic
field inside as $\vec{B}_{in}$. 
Because the component of the field $\vec{H}_{in}=\vec{B}_{in}/\mu$ parallel to the surface ( at $x=0$ ) of the metal is continuous at the surface,
the magnetic field $\vec{B}_{in}$ is obtained such that $\vec{B}_{in}=\vec{B}\mu/\mu_0$ where we denote the vacuum permeability $\mu_0$ ( $\mu_0=1$ in natural unit ).
For instance, $\mu\simeq \mu_0$ for copper or $\mu\sim 5000\times \mu_0$ for iron. 
We should note that in general, the permeability $\mu$ depends on the magnetic field $\vec{B}$ and that
the strength $\vec{B}_{in}=\vec{B}\mu(B)/\mu_0$ saturates as the external field $\vec{B}$ increases.  The maximal magnetic field $B_{in}$ is
at most $1$T$\sim 2$T. Thus, we cannot make $B_{in}$ increase unlimitedly. Hereafter, we assume that $\mu$ is independent of $B$ and we use natural unit $\mu_0=1$.

The electromagnetic fields in the metal are described by the modified Maxwell equations including non trivial permeability of the metal,

\begin{eqnarray}
\label{metal}
\vec{\partial}\cdot\vec{D}+g_{a\gamma\gamma}\vec{\partial}\cdot(a(\vec{x},t)\vec{B}_{in})&=0&, \quad 
\vec{\partial}\times \Big(\vec{H}_{in}-g_{a\gamma\gamma} a(\vec{x},t)\vec{E}\Big)-
\partial_t\Big(\vec{D}+g_{a\gamma\gamma} a(\vec{x},t)\vec{B}_{in}\Big)=\vec{J}_e,  \nonumber  \\
\vec{\partial}\cdot\vec{B}_{in}&=0&, \quad \vec{\partial}\times \vec{E}+\partial_t \vec{B}_{in}=0.
\end{eqnarray}
where  $\vec{H}_{in}=\mu^{-1} \vec{B}_{in}$ and $\vec{D}=\epsilon \vec{E}$ with permittivity $\epsilon$. 
The permittivity $\epsilon$ is nearly equal to $1$ in the metal for radiations
with frequency $\sim1$GHz.
In eq(\ref{metal}) we have included the current $\vec{J}_e=\sigma\vec{E}$ induced by electric field $\vec{E}$, but have neglected external current
generating the background magnetic field $\vec{B}$. 

When we neglect axion contribution, we obtain magnetic field $B^0_{in}=\mu B$ uniform inside the metal where $B$ is external magnetic field imposed.
Obviously, there is no electric field inside the metal.
When we take into account the axion contribution, the oscillating electric field is induced. Naively we expect that the electric field is given such that 
$\vec{E}^{in}_a=-g_{a\gamma\gamma} a(t)\mu\vec{B}/\epsilon$. But, this is not correct as we show below.
 
Assuming the parameter $g_{a\gamma\gamma} a(\vec{x},t)$ small and noting that the electric field is the order of  $g_{a\gamma\gamma} a(\vec{x},t)$,
we derive the equations,

\begin{equation}
\label{max}
\vec{\partial}\cdot \delta\vec{E}=0, \quad \vec{\partial}\times \delta\vec{B}=\mu\vec{J}_e+\mu\epsilon\partial_t\delta\vec{E}, \quad
\vec{\partial}\cdot \delta\vec{B}=0, \quad \mbox{and} \quad \vec{\partial}\times \delta\vec{E}+\partial_t\delta\vec{B}=0 .
\end{equation}
with $\delta\vec{E}=\vec{E}-\vec{E}^{in}_a$ and $\delta\vec{B}\equiv\vec{B}_{in}-\mu \vec{B}=\vec{B}_{in}-\vec{B}_{in}^0$,
where we have used the relation $\vec{\partial}\times \vec{E}^{in}_a=0$ because $\vec{\partial}\times \vec{B}=0$ inside the metal.
Here, we should note that $\delta\vec{B}$ is the order of $g_{a\gamma\gamma} a(\vec{x},t)$.
The electric field $\vec{E}^{in}_a$ is the naive one expected in the metal.

Using the Ohm's law $\vec{J}_e=\sigma\vec{E}$ in eqs(\ref{max}), we derive 
the equation for $\vec{E}$,

\begin{equation}
\label{19}
(\vec{\partial}^2-\mu\epsilon \partial_t^2)\vec{E}=\sigma\mu \partial_t\vec{E}-\mu\epsilon\partial_t^2\vec{E}^{in}_a
\end{equation}
where we note that $\vec{E}^{in}_a\propto \cos(m_at)$. Then, we find the solution,

\begin{equation}
\label{21}
\vec{E}\simeq \vec{E}_0\exp(-\frac{x}{\delta})\cos(\omega t-\frac{x}{\delta})+\frac{\epsilon}{\sigma}\partial_t \vec{E}^{in}_a,
\end{equation} 
with arbitrary field strength $\vec{E}_0$, and frequency $\omega$. The skin depth $\delta$ is given by $\delta=\sqrt{2/\sigma\mu\omega}$.
In the derivation, we have neglected the term $\mu\epsilon \partial_t^2\vec{E}$ in the left hand side of eq(\ref{19}), 
which is much smaller than the term $\sigma\mu \partial_t\vec{E}$
in the right hand side.
Namely, we have used the inequality $\mu\sigma \gg \omega \, ( \sim m_a) $ because the electric conductivity $\sigma \sim 10^4$eV in copper or iron is much larger than 
the axion mass $m_a=10^{-6}$eV $\sim 10^{-4}$eV under consideration.

\vspace{0.1cm}
The first term in the solution $\vec{E}$ represents oscillating electric field with the skin depth $\delta$, while the second term represents
the oscillating electric field $\vec{E}^{in}$ inside the metal; 
$\vec{E}^{in}\equiv \frac{\epsilon}{\sigma}\partial_t \vec{E}^{in}_a$.
The first term is only present in the surface of the metal and represents a radiation with frequency $\omega$ entering from outside the metal.
The strength $\vec{E}_0$ is determined by the boundary condition at the surface $x=0$.
In our case it is determined by electric field induced outside the metal, i.e. electric field in the vacuum. It is 
just $\vec{E}_a=-g_{a\gamma\gamma} a(t)\vec{B}$.
Because of the continuity of the electric field parallel to the surface,
the electric field in the surface is given by $\vec{E}^{suf}=-g_{a\gamma\gamma} a_0\vec{B}\exp(-\frac{x}{\delta})\cos(m_a t-\frac{x}{\delta})$.
( Similar consideration in the superconductor leads to the electric field $\vec{E}_a$ derived in the previous section. )

On the other hand, the electric field $\vec{E}^{in}$ present inside the metal is the one induced by the dark matter axion. 
It is suppressed by the factor $m_a/\sigma$ compared with the naive one $\vec{E}^{in}_a$.
This oscillating electric field
induces the oscillating current $\vec{J}_e=\sigma \vec{E}^{in}= \epsilon\partial_t\vec{E}^{in}$. 

In general, electric field is absent inside conductor with finite size because the field is screened by the electric field produced by surface charge.
It is induced on the surface of the boundary between the conductor and vacuum.
( Even if the electric field is present inside the metal, free electrons move to make the field
cancelled by the surface charge. The surface is perpendicular to 
the electric field. )
In our case there is no such surface charge
because we consider the conductor extended infinitely
in the direction $z$ and $y$. Thus we have non vanishing electric field $\vec{E}^{in}$ inside the metal.

When the conductor has finite size, the electric field $\vec{E}_{in}$ is absent. But we show that
the oscillating electric current $\vec{J}_e=\epsilon \partial_t \vec{E}^{in}_a$ is present.
We suppose that the shape of the metal is cylindrical and the metal has finite size with the length $l$ 
in the direction $z$; the metal has two ends of upper and lower surfaces at $z=0$ and $z=l$ respectively. 
The external magnetic field $\vec{B}$ imposed in the direction $z$ parallel to the cylinder
is perpendicular to the upper and lower surfaces. Then, there is the magnetic field $\mu \vec{B}$ inside the metal.
The magnetic field perpendicular to the surfaces is continuous at the surfaces 
and so it is given by $\mu\vec{B}$ just ouside the surfaces.
Thus, the electric field $\vec{E}$ just outside the upper or lower surfaces induced by the axion is given by
$\vec{E}=-g_{a\gamma\gamma}a(t) \mu\vec{B}=\vec{E}^{in}_a$.
Differently to the case of the magnetic field,
the electric field is absent inside the metal. Thus the electric field perpendicular to the surfaces is discontinuous at the surfaces.
Then, there are surface charges density $\sigma_f$ on the upper and lower surfaces; $\sigma_f=\pm E^{in}_a$.
Because the field $E^{in}_a$ oscillates, the charge density also oscillates. 
It means that an oscillating electric current flowing in the direction $z$ is produced\cite{wire} 
such that $\vec{J}_e=(0,0,J_e)$ with $J_e=\partial_t\sigma_f=\partial_t E^{in}_a$. 
This current only flows through the side surface to the skin depth; the side surface is the one extended in $z$ and $\theta$ directions. 
It is the physical reason why the oscillating current $J_e$ is generated in the metal.
( We have set $\epsilon=1$ in the argument because $\epsilon\simeq 1$ for the electric field with the frequency $m_a/2\pi\sim 1$GHz.  ) 

\vspace{0.1cm}
We make a comment that the electric field $E^{in}= \frac{\epsilon}{\sigma} \partial_t E^{in}_a$ vanishes in the limit of
infinite conductivity, $\sigma \to \infty$. Namely, there is no electric field in perfect conductor.
It means that the above formulation cannot be applied to the superconductor, although the superconductor is perfect conductor.
We have the electric field $\vec{E}_a$ in the surface of the superconductor, as we have shown.
We need a model like Ginzburg-Landau model for the superconductor to appropriately describe electromagnetic fields
in the superconductor.

\vspace{0.2cm}

The current density $J_e=\partial_t E^{in}_a= m_a \mu g_{a\gamma\gamma} a_0\vec{B}\sin(m_at)$ flows through the side surface of the metal, while
there is an additional current $\vec{J}^{suf}_e$ flowing through the surface. It is given such that
$\vec{J}^{suf}_e=\sigma \vec{E}^{suf}=-\sigma g_{a\gamma\gamma} a_0\vec{B} \exp(-\frac{x}{\delta})\cos(m_a t-\frac{x}{\delta})$,
because the electric field $\vec{E}^{suf}$ is present in the surface. Obviously,
the current density $\vec{J}^{suf}_e$ is much larger than  
$J_e$. 
Differently to the superconducting current $\vec{J}_c$, the energy of the current $\vec{J}^{suf}_e$ 
is dissipated; $\int^{2\pi/m_a}_0 dt\, \vec{J}_e^{suf}(t)\cdot \vec{E}^{suf}(t)\neq 0$. 
Because the current oscillates,
it generates dipole radiation from the metal.

\vspace{0.1cm}
In this section we find that electric field $\vec{E}^{suf}=-g_{a\gamma\gamma} a_0\vec{B}\exp(-\frac{x}{\delta})\cos(m_a t-\frac{x}{\delta})$ 
is induced in the surface of the metal, which produces 
the surface current $\vec{J}^{suf}_e=\sigma \vec{E}^{suf}$. The strength of the electric field $\vec{E}^{suf}_a$ is almost identical
to that of the electric field $\vec{E}_a$ in the superconductor; $ \vec{E}^{suf}_a(x=0)=\vec{E}_a$. 

%


\section{radiation flux from cylindrical conductors}
\label{5}

We proceed to show how large amount of radiation is emitted from the superconductor as well as the normal conductor.
The radiation is generated by the oscillating current $\vec{J}_c$ ( $\vec{J}^{suf}_e$ ) in the superconductor ( normal conductor ) induced 
by the oscillating electric fields $\vec{E}_a$ ( $\vec{E}^{suf}_a$ ).
Because the oscillation is harmonic, the radiation is dipole radiation.
The current is carried by Cooper pair ( electrons ) in the superconductor ( normal conductor ).
 
According to the Drude model, we give a simple argument for the form of the current density $J_c=q^2E_an_c/m_am_c$ ( $J^{suf}_e=\sigma E^{suf}_a$ ).
The Cooper pair in the superconductor oscillates with the frequency $m_a/2\pi$
according to the equation of motion, $m_c\dot{v}_c=qE_a\propto \cos(m_at)$; $v_c$ denotes velocity.
We note that the motion of the Cooper pair is not disturbed by impurities in the superconductor,
so there is no dissipative term in the equation of motion.
On the other hand, the electrons in the metal obey the equation of motion,  $m_e\dot{v}_e=eE^{suf}_a-m_e\tau^{-1} v_e$ with the dissipation term,
where $\tau$ denotes relaxation time; when $E^{suf}_a=0$, $v_e(t)\propto \exp(-t/\tau)$.
Obviously, the  second term ( $-m_e\tau^{-1} v_e$ ) in the equation describes the dissipation of electron energy. 
From these equations of motion, we derive the velocity such that 
$v_c=q\int^t dt' E_a(t')/m_c\simeq q E_a/m_am_c$ ( $v_e=eE^{suf}_a\tau/m_e$ ) for the superconductor ( metal ).

Then, the current density in the superconductor is given by $J_c=qn_cv_c=q^2E_a n_c/m_am_c$
with the number density $n_c$ of the Cooper pair ; $n_c\sim 10^{21}\rm cm^{-3}$. 
The formula is identical to the one derived above using the Ginzburg-Landau model.
( We remind that the number density $n_c$ is given such as $n_c=2m_c\langle\Phi\rangle^2=2m_cv_0^2$ in Ginzburg-Landau model. ) 
Similarly, the current density $J_e$ in the metal is given by $J_e=en_e v_e=e^2n_e E^{suf}_a\tau/m_e=\sigma E^{suf}_a$
with the electric conductivity $\sigma\equiv e^2n_e\tau/m_e$.
Both of the currents flow only in the surface of the conductors.

\vspace{0.1cm}
We make a comment that the current $J_c=qn_c v_c=q\int^t dt' E_a(t')/m_c\propto \sin(m_a t)$ has a phase different by $\pi/2$ with 
the electric field $E_a\propto \cos(m_a t)$. It leads to the dissipation less current in the superconductor, as we have shown in previous section.
On the other hand, the current $J_e=en_e v_e=e^2n_e E^{suf}_a\tau/m_e$ has the identical phase to that of the electric field $E^{suf}_a$ so that the current 
in the normal conductor is 
dissipative.

\vspace{0.1cm}
The current oscillates with the frequency $m_a/2\pi$ and flows in the direction of the magnetic field.
The spectrum of the axions has the peak frequency $m_a/2\pi$
with small bandwidth $\Delta\omega\simeq 10^{-6}\times m_a$, because of the velocity $v_a\sim 10^{-3}$ of the dark matter axion in our galaxy. 
Thus, the oscillating current has the same spectrum as that of the axion.

\vspace{0.2cm}
We have proposed a method for the conversion of the dark matter axions to electromagnetic waves.
We use a superconductor of cylindrical shape, on which
the magnetic field $\vec{B}$ parallel to the direction along the length of the superconductor is imposed. 
We take the direction as $z$ direction. The magnetic field is expelled from the superconductor. But
the field penetrates into the superconductor to the depth $\lambda=\sqrt{m_c/q^2n_c}$ ( London penetration depth ).
Thus, the oscillating current is present only in the surface to the depth $\lambda$.
In general, the oscillating current in normal conductors is only present in the surface with the skin depth $\delta=\sqrt{2/\mu\omega\sigma}$.
It also holds for superconductor. But, the skin depth in superconductor is much larger than the penetration depth of magnetic field.
This is because 
the conductivity $\sigma=e^2n_e\tau/m_e$ caused by normal conducting electrons remaining in the superconductor is much small
due to small number density of such electrons. ( The formula $\delta=\sqrt{2/\mu\omega\sigma}$ is derived for normal
conducting current satisfying the Ohm law $J=\sigma E$. The superconducting current does not satisfy the law. )
Typically, the penetration depth is $\lambda\sim 10^{-5}$cm, while the skin depth is $\delta\sim 10^{-4}$cm even for copper
with large conductivity for the frequency $\omega=1$GHz.
Thus, the oscillating current $J_c \propto B$ in the superconductor is only present  to the penetration depth.

%
%

\vspace{0.2cm}
Now, we estimate the radiation flux emitted by the cylindrical superconductor under the magnetic field $B$.
We suppose that the superconductor has radius $R=1$cm and length $l=10$cm.
Then,  the flux of the dipole radiation is given by

\begin{equation}
\label{S}
S_c=\frac{m_a^2 (2\pi Rl\lambda J_c)^2}{3}=\frac{m_a^2(q^2E_0 n_c)^2(2\pi Rl\lambda)^2}{3m_a^2m_c^2}=\frac{(k_a\alpha B)^2(2\pi Rl)^2\rho_a}{3\pi^2m_a^2f_a^2\lambda^2}
\end{equation}
with $E_a\equiv E_0\cos(m_a t)$ ( $E_0=-k_a\alpha a_0B/f_a\pi=-g_{a\gamma\gamma}a_0B$ ),
where we have used the formulae of the penetration depth $\lambda=\sqrt{m_c/q^2n_c}$.

\vspace{0.1cm}
When we use the superconductor Nb$_3$Sn, the penetration depth is about $\lambda=8\times 10^{-6}$ cm
( the number density of Cooper pair $n_c=m_c/q^2\lambda^2\sim 10^{21}\rm cm^{-3}$. )
Then, we numerically estimate the flux $S$,

\begin{equation}
\label{NS}
S_c\simeq 4.4\times 10^{-18}\mbox{W}\Big(\frac{8\times10^{-6}\rm cm}{\lambda}\Big)^2\Big(\frac{B}{5\rm T}\Big)^2
\Big(\frac{R}{1\rm cm}\Big)^2\Big(\frac{l}{10\rm cm}\Big)^2\Big(\frac{k_a}{1.0}\Big)^2\Big(\frac{\rho_a}{0.3\rm GeV/\rm cm^3}\Big),
\end{equation}
with $k_a\simeq 0.37$ for DFSZ model and $k_a\simeq -0.96$ for KSVZ model,
where there is no dependence on the axion mass.
The spectrum of the radiation shows a sharp peak at the frequency $m_a/2\pi$ with the bandwidth $\Delta \omega\sim 10^{-6}m_a/2\pi$.
It is remarkable that the radiation flux in eq(\ref{NS}) is four times of the order of magnitude larger than that obtained in resonant cavity\cite{admx,sikivie}.

The flux is obtained by integrating a Poynting vector over the sphere with radius $r\gg 2\pi/m_a$ 
around the superconductor; $S_c=\int S_c(\theta, r) r^2 d\Omega=\int S_c(\theta, r)r^2 \sin\theta d\theta d\phi$,
where 

\begin{equation}
S_c(\theta, r)=\frac{m_a^2 (2\pi Rl\lambda J)^2(\sin\theta)^2}{8\pi r^2} ,
\end{equation}  
where we have taken the polar coordinate.

The dipole radiation is emitted mainly toward the direction ( $\theta=\pi/2$ ) perpendicular to the electric current flowing in $z$ direction.
Thus, when we measure the radiation emitted
in the direction, we receive relatively strong flux density.  
For example, when we observe the radiation using the radio telescope of parabolic dish antenna with the diameter $32$m by putting the cylindrical 
superconductor $100$m away from the telescope
( e.g. Yamaguchi 32-m radio telescope of National Astrophysical  Observatory of Japan ),  
the observed flux per frequency $P_c$ is given by

\begin{equation}
P_c\equiv \int \frac{S_c(\theta, r)}{\Delta \omega} r^2 d\Omega=\int^{\pi/2-\delta_t}_{\pi/2+\delta_t} \frac{S_c(\theta, r)}{\Delta \omega} r^2\sin\theta d\theta d\phi\simeq 
\frac{m_a^2 (2\pi Rl\lambda J)^2\delta_t^2}{8\Delta \omega}=\frac{3S_c\delta_t^2}{8\Delta \omega}\simeq 0.4\times 10^{-22}\rm W/Hz,
\end{equation}
with $\delta_t \simeq 16\rm m/100\rm m=0.16$ and $\Delta\omega=10^3$Hz$\big(m_a/(6\times 10^{-6}\rm eV)\big)$,
where the center of the parabolic antenna is set in the direction $\theta=\pi/2$.  
Thus, we find that the antenna temperature is approximately $T_a\equiv P\eta\simeq 1.5$K with the unit $k_B=1$,
assuming the antenna efficiency $\eta \simeq 0.6$. Therefore, the radiation can be observed
with the radio telescope with diameter such as $32$m.

\vspace{0.1cm}
We estimate the detection sensitivity. When we observe the radiation over time $t$ with bandwidth $\delta \omega$, 
the ratio of signal to noise is given by $S/N=\frac{\bar{S}_c}{T_{sys}}\sqrt{t/\delta \omega}$ where $\bar{S}_c=3\delta_t^2S_c/8\sim 10^{-2}S_c$ denotes
the radiation flux received by the telescope $100$m away from the superconductor and
$T_{sys}$ is the system noise temperature. 
For instance, $T_{sys}=40$K for Yamaguchi 32-m radio telescope.
Therefore, we find that

\begin{equation}
\label{sen}
\frac{S}{N}\sim 40\times \sqrt{\frac{1\rm MHz}{\delta \omega}}\sqrt{\frac{t}{1\,\rm s}}
\Big(\frac{g_{15}}{m_6}\Big)^2\Big(\frac{5\times10^{-6}\rm cm}{\lambda}\Big)^2\Big(\frac{B}{7\rm T}\Big)^2
\Big(\frac{R}{2\rm cm}\Big)^2\Big(\frac{l}{20\rm cm}\Big)^2\Big(\frac{\rho_a}{0.3\rm GeV/\rm cm^3}\Big)
\end{equation}
with $T_{sys}=40$K, where we have taken the physical parameters $B=7$T, $R=2$cm and $l=20$cm to have 
better detection sensitivity. Here, we put $g_{15}\equiv g_{a\gamma\gamma}/(10^{-15}\rm GeV^{-1})$ and  $m_6\equiv m_a/(10^{-6}\rm eV)$.
Thus, even for DFSZ axion ( $(\frac{g_{15}}{m_6})^2\simeq 0.1$ ), we reach the sensitivity $S/N \sim 4$ 
when we observe the radiation over $1$ second 
with the bandwidth $\delta\omega=1$MHz. In the formula we do not use the relation $m_af_a\simeq 6\times 10^{-6}\rm eV\times 10^{12}$GeV 
specific to the QCD axion. 
The formula holds even for axion-like particle.

Obviously, the larger radiation flux can be obtained when
we put the superconductor nearer the radio telescope than $100$m.  
Furthermore, even when we use a radio telescope with smaller radius than $32$m, 
large $S/N$ ratio can be achieved if we put the apparatus near the telescope.  
The radiation flux is determined by the solid angle of the parabolic dish antenna viewed from the superconductor. 
The larger solid angle leads to larger radiation flux.
The merit of our proposal is that we can simultaneously search wide bandwidth of the radio frequency without tuning 
the shape of the superconductor. 
In this way, we can observe the radiation from the dark matter axion.  

\vspace{0.1cm}
We make a comment on the actual setup for the observation.
The strong magnetic field $B$ parallel to the direction along the length of the cylindrical superconductor is produced by coils surrounding it.
The coils should have open space for the dipole radiations to escape outside the coils and reach the telescope. In particular, the open space should be present
in the $\theta=\pi/2\pm \delta $ directions ( e.g. $\delta\simeq 16\rm cm/100\rm cm=0.16$ )  perpendicular to the cylindrical superconductor. 
That is, the coils are composed of two parts; one covers the upper side of the superconductor and the other one covers the lower side.
Then, there is an open space in the coils through which the radiations can escape from the coils.
Furthermore,
the whole of the apparatus must be cooled. We need to use a glass container for liquid helium so as for the radiation to pass the container and
reach radio telescope.

\vspace{0.1cm}
We make an additional comment on radiation from magnetic vortex.
We use type 2 superconductor for the apparatus in order for the strong magnetic field not to break the superconductivity.
Then, the magnetic vortices are formed inside the superconductor and the electric field $E_a$ is induced around the vortices.
But, the radiations from magnetic vortices 
do not arise. This is owing to the fact that
each vortex is surrounded by the superconducting state $\langle \Phi \rangle \neq 0$.
The electromagnetic waves do not pass the state.

\vspace{0.2cm}

For comparison,
we estimate the radiation flux $S_e$ from the normal conductor with the shape identical to the one in the superconductor. 

\begin{eqnarray}
\label{N}
S_e&=&\frac{m_a^2 (2\pi Rl\delta J_e)^2}{3}=\frac{m_a^2(\sigma E_0 )^2(2\pi Rl\delta)^2}{3}=\frac{4(k_a\alpha B)^2(2\pi Rl)^2\rho_a}{3\pi^2m_a^2f_a^2\delta^2} \\
&\simeq&2.8\times 10^{-20}\mbox{W}\Big(\frac{2\times10^{-4}\rm cm}{\delta}\Big)^2\Big(\frac{m_a}{4\times 10^{-6}\rm eV}\Big)\Big(\frac{B}{5\rm T}\Big)^2
\Big(\frac{R}{1\rm cm}\Big)^2\Big(\frac{l}{10\rm cm}\Big)^2\Big(\frac{k_a}{1.0}\Big)^2\Big(\frac{\rho_a}{0.3\rm GeV/\rm cm^3}\Big),
\end{eqnarray}
with $E_0=g_{a\gamma\gamma} a_0\mu B$ and $\delta=\sqrt{2/m_a\sigma}$,
where we put $\mu=1$ for simplicity.

Differently to the case of the superconductor, the radiation flux depends on the axion mass.
This is because the skin depth $\delta$ depends on the frequency of the radiation. 
The difference between the flux from the superconductor and that from the normal conductor
comes from the difference in the depth in which the currents flow.
This difference causes the big difference in the flux.
That is,  
the radiation flux $S_e$ emitted from the normal conductor is about a hundred times smaller than $S_c$ from the superconductor
because typically $\delta\sim 10^{-4}$cm and $\lambda\sim 10^{-5}$cm.

\section{superconducting resonant cavity}
\label{5.5}

We would like to show radiation energy generated in superconducting resonant cavity,
comparing it with energy in normal conducting cavity. 
The cylindrical resonant cavity ( tube ) has been considered for the detection of the dark matter axion for more than 30 years ago\cite{1sikivie}.
The cavity is formed of normal conductor, e.g. cupper.
In this section we consider superconducting resonant cavity and show that a hundred times large amount of energy
is produced inside it, compared with that in the normal conductor.
 In order to do so, it is instructive to note
the difference between the superconductor and the normal conductor.
The difference is the currents flowing through the conductors. 
That is, the electric currents $J_c= E_a/(m_a\lambda^2)$ flows through the surface of the superconductor 
and $J_e=2E^{suf}_a/(m_a\delta^2)$ does through the surface of the metal. 
( We remember $E_a\simeq E^{suf}_a $ and
$J_c =q^2n_c E_a/(m_a m_c)$ ( $J_e=\sigma E^{suf}_a$ ) with $\lambda^{-2}=q^2n_c/m_c$ ( $\delta^{-2}=\mu m_a\sigma/2$ ).
We also note the difference in the depth of the current flow, which leads to the difference in their fluxes produced by
the cylindrical conductors 
as shown in the previous section.

\vspace{0.1cm}
In the derivation\cite{sikivie,kim} of the radiation energy inside the cavity, 
we solve equations of electromagnetic fields coupled with the axion by taking account of
the effect of the normal conducting cavity. The effect is taken by introducing the current $J_e=\sigma E_a\simeq 2E_a/(m_a\delta^2) $ in the equations.

We use cylindrical coordinate $(\rho, \theta, z)$ in which the
the cavity is located at $\rho=R_c$. The magnetic field $B$ is imposed in the direction $z$ parallel to the cavity.
It is assumed to be uniform in the cavity.
We solve the equations for the TM mode of the electric field $\vec{E}=(0,0,E_z)$ and the magnetic field $\vec{B}=(0,B_{\theta},0)$ 
produced by the dark matter axion.  
The equations inside the normal conducting cavity ( $\rho<R_c$ ) are identical to the ones
inside the superconducting cavity. The equations hold in vacuum with magnetic field imposed.
Assuming the cylindrical symmetry, the equations are in the following,

\begin{equation}
(\partial_{\rho}^2+\frac{1}{\rho}\partial_{\rho}-\partial_t^2)E_z^{in}-g_{a\gamma\gamma}\partial_t^2a(t)B=0
\quad \mbox{and} \quad \partial_{\rho} E_z^{in}=\partial_t B_{\theta}^{in}
\end{equation}

The solutions describing the magnetic and electric fields in both cavity are given by

\begin{equation}
\label{cav}
B_{\theta}^{in}=b(t) J_1(m_a\rho) \quad \mbox{and} \quad E_z^{in}=-\frac{\partial_t b(t) J_0(m_a\rho)}{m_a}-g B a(t)
\quad \mbox{for} \quad \rho<R_c,
\end{equation}
where $J_{0,1}(m_a\rho)$ denotes Bessel functions of the first kind. The function $b(t)$ is determined by the 
boundary conditions at $\rho=R_c$. Thus, the function $b(t)$ takes different values depending on
the normal conductor or superconductor. 

\vspace{0.1cm}
In the normal conductor, by solving both equation (\ref{19}) and equation $\partial_{\rho} E_z=\partial_t B_{\theta}$ 
with the cylindrical coordinate,
we find electric and magnetic fields in the region $\rho \ge R_c$

\begin{eqnarray}
B_{\theta,e}^{out}&=&\Big(b_0^r\cos(\frac{\rho-R_c}{\delta}-m_a t)-b_0^i\sin(\frac{\rho-R_c}{\delta}-m_at)\Big) 
\exp\Big(-\frac{(\rho-R_c)}{\delta}\Big) \\
E_{z,e}^{out}&=&-\frac{m_a\delta}{2}\Big((b_0^r+b_0^i)\cos(m_a t-\frac{\rho-R_c}{\delta})
-(b_0^r-b_0^i)\sin(m_a t-\frac{\rho-R_c}{\delta})\Big)\exp\Big(-\frac{(\rho-R_c)}{\delta}\Big) 
\end{eqnarray}
with the coefficients $b_0^r$ and $b_0^i$,

\begin{equation}
b_0^r=\frac{m_a\delta}{2}\frac{-g_{a\gamma\gamma} a_0 B}{Z^2+(\frac{m_a\delta}{2})^2} 
\quad \mbox{and} \quad b_0^i=\frac{g_{a\gamma\gamma} a_0 B Z}{Z^2+(\frac{m_a\delta}{2})^2}
\quad \mbox{with} \quad Z=\frac{J_0(m_aR_c)}{\mu J_1(m_aR_c)}-\frac{m_a\delta}{2} 
\end{equation}
where we have neglected the small term $m_a/\sigma$ in the solution $E_{z,e}^{out}$ which is present in the solution eq(\ref{21}).

These coefficients are found by imposing the boundary conditions $B_{\theta,e}^{in}=\mu B_{\theta,e}^{out}$ and
$E_{z,e}^{in}=E_{z,e}^{out}$ at $\rho=R_c$, where the fields $B_{\theta,e}^{in}$ and $E_{z,e}^{in}$ in the region $\rho \le R_c$ are given by 
 
\begin{eqnarray}
 B_{\theta,e}^{in}&=&\frac{b_0^r\cos(m_at)+b_0^i\sin(m_at)}{\mu J_1(m_aR_c)}J_1(m_a\rho) \\
 E_{z,e}^{in}&=&\frac{b_0^r\sin(m_at)-b_0^i\cos(m_at)}{\mu J_1(m_aR_c)}J_0(m_a\rho)-g_{a\gamma\gamma}a(t)B .
 \end{eqnarray} 
with $b(t)=(b_0^r\cos(m_at)+b_0^i\sin(m_at))/\mu J_1(m_aR_c)$ in eq(\ref{cav}).

We find that the electromagnetic fields $B_{\theta,e}^{out}$ and $E_{z,e}^{out}$ in the normal conductor
are only present in the surface to the depth $\delta$ owing to the skin effect. 

\vspace{0.1cm}
On the other hand, in the superconductor, 
by solving  both equation (\ref{delta3}) and equation $\partial_{\rho} E_z=\partial_t B_{\theta}$ 
we find electric and magnetic fields in the region $\rho \ge R_c$

\begin{eqnarray}
B_{\theta,c}^{out}&=&\Big(\frac{1}{m_a\lambda}A_s
+\frac{g_{a\gamma\gamma} a_0B}{m_a\lambda}\Big)\sin(m_at)\exp\Big(-\frac{(\rho-R_c)}{\lambda}\Big)   \\
E_{z,c}^{out}&=&-\Big(A_s
+g_{a\gamma\gamma} a_0B\Big)\cos(m_at)\exp\Big(-\frac{(\rho-R_c)}{\lambda}\Big) 
\end{eqnarray}
with $A_s=g_{a\gamma\gamma} a_0B J_0(m_a R_c)/(m_a\lambda J_1(m_aR_c)-J_0(m_aR_c)) $.

The fields satisfy the boundary conditions $B_{\theta,c}^{out}=B_{\theta,c}^{in}$ and $E_{z,c}^{out}=E_{z,c}^{in}$ at $\rho=R_c$,
where $B_{\theta,c}^{in}$ and $E_{z,c}^{in}$ in the region $\rho \le R_c$ are given by 

\begin{eqnarray}
 B_{\theta,c}^{in}&=&\frac{g_{a\gamma\gamma} a_0 B\sin(m_at)}{m_a\lambda J_1(m_aR_c)-J_0(m_aR_c)}J_1(m_a\rho) \\
 E_{z,c}^{in}&=&\frac{-g_{a\gamma\gamma} a_0 B\cos(m_at)}{m_a\lambda J_1(m_aR_c)-J_0(m_aR_c)}J_0(m_a\rho)-g_{a\gamma\gamma}a(t)B .
 \end{eqnarray} 
with $b(t)=(A_s+g_{a\gamma\gamma} a_0 B)\sin(m_at)/m_a\lambda J_1(m_aR_c)$ in eq(\ref{cav}). 
We find that the electromagnetic fields $B_{\theta,c}^{out}$ and $E_{z,c}^{out}$ in the superconductor
are only present in the surface to the depth $\lambda$ owing to the Meissner effect.

\vspace{0.1cm}
Because we have obtained the electric and magnetic fields inside the cavity ( $\rho<R_c$ ), we estimate their energies
$U_e=\int dV \langle((E^{in}_z)^2+(B^{in}_{\theta})^2)\rangle/2$ where
the time average is taken; $\frac{m_a}{2\pi}\int_0^{2\pi/m_a} dt \,Q=\langle Q\rangle$.
In the estimation we use the resonant condition $J_0(m_a R_c)=0$, which enhances the 
field strengths in the cavity; the value $m_aR_c$ is the first zero point of the Bessel function $J_0(\rho)$.
Indeed, it is easy to see that when we put $J_0(m_a R_c)=0$,
the field strengths are enhanced because the factor $b_0^{r,i}\propto (m_a\delta)^{-1} (\,\, \gg 1$ )
of the fields in the normal conductor are large and 
the factor $\big(m_a\lambda J_1(m_aR_c)-J_0(m_aR_c)\big)^{-1}=\big(m_a\lambda J_1(m_aR_c)\big)^{-1} ( \,\, \gg 1 $ ) of
the fields inside the superconductor are also large. On the other hand, when $J_0(m_a R_c)$ is of the order of unity,
the fields are not enhanced. Actually,
the factor $b_0^{r}\propto m_a\delta $ is small, but the factor $b_0^1\propto Z^{-1}$ is of the order of unity, while
the factor $m_a\lambda J_1(m_aR_c)-J_0(m_aR_c)$ 
is of the order of unity. 

\vspace{0.1cm}
Using $\mu\simeq 1$ for copper, we find that the field energies $U_e$ in the normal conducting 
and $U_c$ in the superconducting cavities are given by, respectively,

\begin{eqnarray}
U_e&=&\frac{2(g_{a\gamma\gamma} B)^2\bar{\rho}_a}{m_a^4\delta^2}V \quad \mbox{for normal conducting cavity} \\
U_c&=&\frac{(g_{a\gamma\gamma} B)^2\bar{\rho}_a}{m_a^4\lambda^2}V \quad \mbox{for superconducting cavity}
\end{eqnarray}
with the volume $V$ of the cavity,
where we denote the density $\bar{\rho}_a$ of the dark matter axion; $\bar{\rho}_a=m_a^2a_0^2/2$.

\vspace{0.1cm}
We find that the difference in the energy between the normal conductor and the superconductor is 
in the depth from surface to which current flows in the conductors. 
That is, the skin depth $\delta$ in the normal conductor and the penetration depth $\lambda$ in the superconductor.
As we have mentioned in section \ref{5}, typically, $\delta\sim 10^{-4}$cm and $\lambda\sim 10^{-5}$cm.
Therefore, a hundred times larger amount of the radiation energy is produced inside the superconducting resonant cavity
than the energy inside the normal conducting cavity.

\vspace{0.1cm}
Contrary to the large amount of the radiation energy produced inside the cavity, 
the flux absorbed in the superconducting cavity vanishes because the time average of the Poynting vector
vanishes at the surface $\rho=R_c$ of the cavity; $\int_0^{2\pi/\omega_a}dt \,\delta \vec{E}\times \delta \vec{B}=0$.
This is owing to the fact that the superconducting current is dissipationless. To find the radiation flux absorbed in the cavity we need to take
into account the effect of electrons remaining in the superconductor
without forming Cooper pairs at nonzero temperature. The electrons absorb the radiation.

\section{summary and discussion}
\label{6}

We have shown that electric field induced by axion in superconductor is essentially identical to the one in vacuum.
The electric field is proportional to magnetic field in the superconductor. 
The electric field is only present in the surface of the superconductor because of the Meissner effect;
magnetic field is expelled from the superconductor.
The result is obtained by analyzing equations of electromagnetic fields coupled with the axion and Cooper
pairs. The Cooper pairs are described by a Ginzburg Landau model.

On the other hand, although the magnetic field is present inside normal conductor,
the electric field induced by axion is absent in normal conductor.
It is only present in the surface of the conductor because of skin effect of the oscillating electric field.
The strength of the electric field is almost equal to the one in vacuum.
The result is obtained by analyzing equations of electromagnetic fields in the metal coupled with the axion.

These electric fields oscillate with the frequency given by the axion mass so that 
Cooper pairs ( electrons ) in the superconductor ( normal conductor ) are forced to oscillate
and emit radiations with the frequency. 
We have estimated the radiation fluxes from the cylindrical conductors. 
In particular, the flux from the superconductor is sufficiently large to be observed
by existing radio telescopes. For instance, even when the superconductor with 
radius $1$cm and length $10$cm is put $100$m away from the radio telescope
of the parabolic dish antenna with the diameter $32$m,
the flux received by the telescope
is four times of the order of the magnitude larger than that in the resonant cavity recently used\cite{admx} in ADMX. 

The flux from the superconductor or metal is inversely proportional to the square of the penetration depth $\lambda$ in the superconductor or the skin depth $\delta$
in the normal conductor, respectively.
In general, the penetration depth $\sim 10^{-5}$ cm is shorter than the skin depth $\sim 10^{-4}$ cm for frequency $\sim 1$GHz.
This difference results in the difference of each flux.

\vspace{0.1cm}
In this paper we have mainly considered the QCD axion whose mass is expected in the range from $10^{-6}$eV $\sim 10^{-4}$eV.
This expectation comes from the previous our paper\cite{iwazakifrb2}, in which we have predicted the axion mass $\sim 7\times 10^{-6}$eV.
The prediction comes from the analysis of the spectrum of fast radio bursts ( FRBs )\cite{frb,frb121102}.
The FRBs are radio bursts with typical frequency $1$GHz, flux $\sim 10^{40}$erg/s and duration $\sim 1$ms. It is still a mysterious phenomena 
in astrophysics.
Our model\cite{iwazakifrb} for the FRBs is that the fast radio bursts arise from the collision between axion star\cite{axionstar} and magnetized dense electron gases such as 
neutron star or geometrically thin accretion disk around black hole with larger mass than $\sim10^3M_{\odot}$. 
The axion star is gravitationally bound state of axions, which is more dense than the dark matter axion $\rho_a$ under the consideration.
The emission mechanism of radiations from the astrophysical objects is identical to the one discussed in the present paper.
That is, the strongly magnetized electron gas emit radiations when they collide the dense axions.
For this reason, our main interest in the axion mass is in the range mentioned above.

Obviously, our method for the detection of the dark matter axion can be applicable for much wide mass range beyond 
the range $10^{-6}$eV $\sim 10^{-4}$eV. We need sensitive receiver for the capture of the radiations from the cylindrical superconductor.
The receiver should have surface area with the large solid angle as possible viewed from the superconductor.
We also need to fabricate appropriate magnet, and glass container for liquid helium, in order for the radiations
to pass through the magnet and the container and to reach the receiver. 
Then, we can search the wide range of the axion mass with high sensitivity.

\vspace{0.2cm}
The author
expresses thanks to Alexander John Miller, Izumi Tsutsui and Osamu Morimatsu  for useful comments and discussions.
Especially, he expresses great thank to Yasuhiro Kishimoto for useful comments.
This work was supported in part by Grant-in-Aid for Scientific Research ( KAKENHI ), No.19K03832.



\end{document}